# Test Generation and Scheduling for a Hybrid BIST Considering Test Time and Power Constraint

Elaheh Sadredini, Mohammad Hashem Haghbayan, Mahmood Fathy, and Zainalabedin Navabi
es9bt@virginia.edu, hashem@cad.ut.ac.ir, mahfathy@iust.ac.ir, navabi@cad.ut.ac.ir

*Abstract*— **This paper presents a novel approach for test generation and test scheduling for multi-clock domain SoCs. A concurrent hybrid BIST architecture is proposed for testing cores. Furthermore, a heuristic for selecting cores to be tested concurrently and order of applying test patterns is proposed. Experimental results show that the proposed heuristics give us an optimized method for multi clock domain SoC testing in comparison with the previous works.**

## I. INTRODUCTION

By increases in complexity of system on chip (SoC) designs, user defined logics and the number of cores in SoCs, achieving a reasonable test time becomes very important. Therefore, different approaches have been used to reduce the total SoC test application time. One approach is based on allocating optimal test access mechanisms (TAMs) and test scheduling algorithms for testing cores [1-4][14][23]. One of the most important factors to improve test application time (TAT) in SoC testing is a suitable test scheduling algorithm based on the appropriate test architectures [4-7][24]. In [7, 11, 26] the authors proposed an algorithm based on algebra and genetic algorithm to find an optimal test scheduling. On the other hand, because of the increasing switching activities during the test process, power consumption of a chip while being tested, is higher than the normal mode of operation and that can affect the reliability of testing [8]. Thus, power constraint and thermal aware SoC testing have become important during concurrent testing [9, 10, 12, 13, 25].

Many test scheduling schemes and algorithms have been proposed to minimize SoC test time utilizing concurrent approaches with power limitation consideration. In [14, 15, 27] the authors presented a heuristic method to minimize test application time by allocating optimum wires of TAM to each core and best test scheduling scheme. In [16, 20] the authors presented a power aware bus architecture and an appropriate algorithm to use a functional bus instead of TAM to test cores concurrently. The main idea is to use memory buffers for applying deterministic test patterns (DTPs) to cores concurrently when the speed of the functional bus is higher than the speed of injecting test patterns in cores. In [21] an architecture and a test scheduling method for testing cores with multi-clock domain SoC is presented. On the other hand, when talking about concurrency, built-in self-test (BIST) becomes a beneficial method to achieve concurrency. In [17] the authors proposed an algorithm to combine BIST with external testing (called hybrid BIST or CBET) to achieve optimum result for test time minimization problem. In [10] the authors proposed an algorithm based on solving rectangular packing problem for power constrained concurrent hybrid BIST for SoC testing. In this paper we propose a hybrid BIST structure and concurrent test scheduling for multi-clock domain SoC. Furthermore a novel algorithm for finding the optimal number of pseudo random test patterns (PRTPs) and deterministic test patterns for each core is presented.

Section II is devoted to discuss test generation process in hybrid BIST. A hybrid BIST architecture for a multi-clock domain SoC is proposed in Section III. In Section IV, a test scheduling graph for modeling power aware core testing is presented. Based on the proposed test scheduling graph, some heuristics for selecting the set of cores to be tested concurrently, determining sets and the order of applying deterministic and pseudo random test patterns to each core will be discussed in Section V. An algorithm for test scheduling based on the test scheduling graph is proposed in Section VI. The results obtained by the proposed methods are drawn in Section VII. Finally Section VIII concludes our work.

## II. CALCULATING NUMBER OF PSEUDO RANDOM TESTS

This section discusses an approach for calculating pseudo random tests to be applied to the hybrid BIST. As mentioned, this BIST uses DTPs (Deterministic Test Patterns) and PRTPs (Pseudo Random Test Patterns). Both DTPs and PRTPs have advantages and disadvantages. DTPs are generated for random-resistant faults and increase fault coverage more than PRTPs. But, in many cases, an automatic test equipment (ATE) is needed for applying test patterns with global clock and through scan chains. So, using DTPs increase the TAT. On the other hand, PRTP can be produced through BIST architectures with local clocks [17]. In some BIST architectures, scan chain is not needed that reduces the number of cycles for applying test patterns. In general, the speed of PRTPs application (through a BIST architecture) is higher than the speed of DTPs (through an external ATE). We refer to $S_{Bi}=F_{Bi}/AC_{Bi}$ and $S_{Ei}=F_{Ei}/AC_{Ei}$ as *BIST speed* and *external speed* of core $i$ respectively. Here $F$ is the clock frequency and $AC$ is the number of cycles for applying each test pattern.

In a hybrid BIST method, both the speed of PRTPs and quality of DTPs will be used for achieving the minimum TAT. In the common hybrid BIST test generation, first, PRTPs are generated for detecting *easy-to-detect* faults and then, for the remaining faults (*random-pattern-resistant faults),* DTPs will be generated. In the end of the test generation process, DTPs are generated for specific faults. Quality of these vectors is very high, and for achieving full coverage, we should apply these test vectors to the CUT. With applying these vectors in the start of test generation, a large number of faults will be detected, which leave a relatively few fault for the PRTPs to detect. Thus reducing the number of PRTPs significantly. In our method a PRTP test generation is sandwiched between two DTPs as discussed below. Generally, there are 3 stages for the test generation process.

The first stage of test generation is phase 1 of deterministic test generation (*DTPs-phase1*). Such test vectors are generated for *very-hard-to-detect* faults and should be applied early in the test process. A Large number of faults (*easy-to-detect* and *hard-to-detect*) will be detected with these high quality tests.

The second stage of test generation is the pseudo-random one. *PRTPs* are generated by linear feedback shift registers (LFSR) and detect the remaining *easy-to-detect* faults. The problem is finding the optimal number of PRTPs for achieving a minimum TAT for the remaining faults. A simple binary search algorithm helps us find the optimal number of PRTPs (Figure 1). This algorithm will be discussed later.

The third stage of test generation is phase 2 of deterministic test generation (*DTPs-phase2*). Such tests are generated for remaining faults. Finally, full coverage will be achieved.

The total TAT decreases by applying the above three phases of test generation process, compared with hybrid BIST tests where deterministic tests follow pseudo random tests. Figure 2 shows test application time for the proposed test generation method (checkered-dark-hatched bars) and work presented in [22] (gray-white bars). In our proposed method, PRTP_PM is the optimal number of PRTPs and PRTP_PW is the optimal number of PRTPs for previous work of [22].

```
Find optimal N_PRTP
1 Inputs: max N_PRTP, min N_PRTP
2 IF TAT (max N_PRTP) < TAT (min N_PRTP)
3     Report max N_PRTP as optimal N_PRTP;
4     Return;
5 END IF
6 T1 = Calculate TAT for min N_PRTP;
7 T2 = Calculate TAT for (max N_PRTP - min N_PRTP)/4;
8 T3 = Calculate TAT for (max N_PRTP - min N_PRTP)/2;
9 T4 = Calculate TAT for 3×(max N_PRTP - min N_PRTP)/4;
10 T5 = Calculate TAT for max N_PRTP;
   // the two above calculation will be used for
   // evaluating the following IF statement.
11 IF T5 > T4 > T3
12     Find optimal N_PRTP between
       Min N_PRTP and (max N_PRTP - min N_PRTP)/2;
13 ELSE
14     Find optimal N_PRTP between
       (max N_PRTP - min N_PRTP)/2 and max N_PRTP;
15 END IF
   END
```

Figure 1. Finding optimal $N_{PRTP}$ in hybrid BIST.

*Calculation of TAT in hybrid BIST:* The overall TAT, is the addition of PRTPs application time and DTPs (phase 1 and phase 2) application time.

$N_{PRTP}$ shows the number of PRTPs and $N_{DTP}$ shows the number of DTPs.

*Finding optimal $N_{PRTP}$ in hybrid BIST:* The process of finding the optimal number of PRTPs is illustrated in Figure 1. The function shown in this figure starts finding the optimal $N_{PRTP}$ between zero $N_{PRTP}$ and a maximum $N_{PRTP}$ as minimum and maximum, respectively (inputs of the function, Line 1). Then, TAT for *min* $N_{PRTP}$ and *max* $N_{PRTP}$ will be calculated. After that, TAT for three $N_{PRTP}$ between *min* and *max* $N_{PRTP}$ will be calculated (Line 6-10) to determine the optimal $N_{PRTP}$ is at the right of the (*max* $N_{PRTP}$ − *min* $N_{PRTP}$)/2 or left. Then the same process continues by assigning the middle place as the *min* or *max* $N_{PRTP}$. This process continues until finding the optimal $N_{PRTP}$. The algorithm is very similar to binary search for finding the minimum TAT. The order of the algorithm is:

$$O(\log_2 n) * (O(Fault\ Simulation) + (Test\ Generation)),$$

Where *n* is the maximum $N_{PRTP}$ used at the start of the algorithm.

### III. CONCURRENT HYBRID BIST IN SOCS

The architecture of concurrent hybrid BIST, used in this paper, is shown in 2.

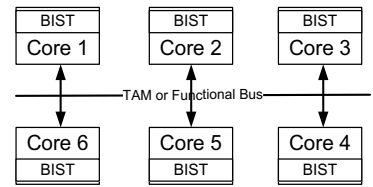

Figure 3. Concurrent Hybrid BIST architecture.

Each core in the SoC shown in Figure 3 has a BIST architecture and a TAM or a functional bus ([16, 20]) which can be used to apply external test patterns. We use this bus for the deterministic tests, while pseudo random tests are generated internal to the core. If we use $vp_{ki}$ for the k$^{th}$ set of PRTPs for the *built-in self-test part* of core *i*, and $vd_{ki}$ for the k$^{th}$ set of DTPs for the *external test part* of core *i*, then:

$$v_{ki} = vp_{ki} \cup vd_{ki} \qquad (3.1)$$
$$F_{v_{ki}} = F_{vp_{ki}} \cup F_{vd_{ki}} \qquad (3.2)$$
$$T(v_{ki}) = T(vp_{ki}) + T(vd_{ki}) \quad (3.3)$$

In the above equations, $v_{ki}$ is the set of test patterns that is applied to core *i*, $F_{v_{ki}}$ is the set of detected faults by $v_{ki}$. $T(v_{ki})$ is the test time that remains for application of remaining vectors in the test set for core *i*. Thus, in the above, the remaining test time is what remains of PRTPs and DTPs.

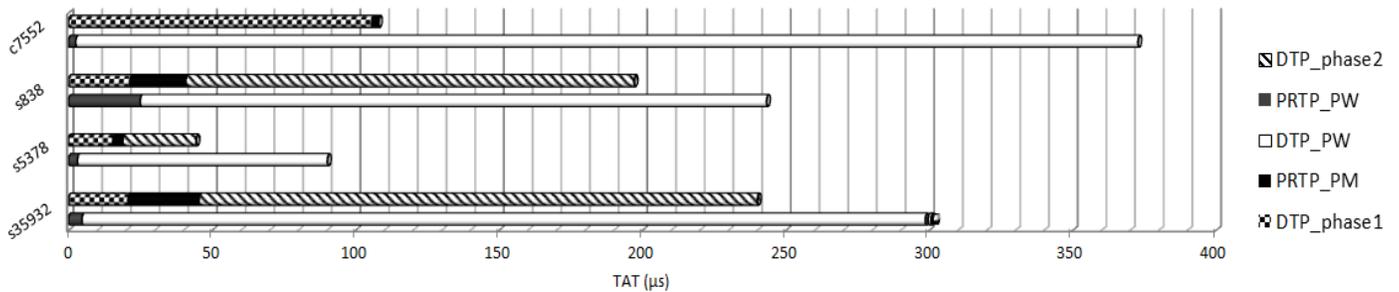

Figure 2. TAT for proposed method (PM) and work presented in [22].

## IV. PEAK POWER LIMITATION AND TEST SCHEDULING

Many works have been done on power constraint SoC test scheduling and, also SoC hybrid BIST. The main contribution in this paper regarding to previous works is proposing a new test generation algorithm for a hybrid BIST architecture, and a test scheduling algorithm based on the generated test patterns. This section presents an algorithm for selecting cores that can be tested concurrently, and a test scheduling graph.

As mentioned before, in a hybrid BIST architecture, the total peak power limitation should be considered while cores are being tested concurrently. It is obvious that power consumption varies over time, but to simplify the test scheduling process, it is assumed that power consumption of each core is the same as its peak power consumption all over the test process [19]. For the rest of the paper, peak power consumption of core $i$ is represented by $P_{mi}$, and $P_{max}$ is the maximum power limit of the SoC.

For presenting the test scheduling algorithm, example of five cores shown in Table 1. This table shows the characteristics of five cores that needs to be available at the start of the algorithm that calculates the set of cores that can be tested simultaneously. The parameters of this algorithm are: peak power consumption of core $i$ ($P_{mi}$), time for applying DTPs ($T(v_d)$), and total time for applying PRTPs ($T(v_p)$).

For example Core 1 in Table 1 has 100uw test peak power, needs 300us for applying all deterministic test patterns using full TAM wires, and needs 200us for applying its PRTPs through BIST architecture. We will use this example for some definitions.

Table 1. Characteristics of the cores for an SoC example

| Core | 1 | 2 | 3 | 4 | 5 |
|---|---|---|---|---|---|
| $P_m$ | 100 | 200 | 50 | 200 | 50 |
| $T(v_d)$ | 300 | 400 | 500 | 150 | 100 |
| $T(v_p)$ | 200 | 500 | 200 | 600 | 300 |

For applying the proposed algorithm, we propose a test scheduling graph model in this section. The following definitions are used in presentation of the scheduling graph.

***Definition 1*** *(power group sets or $N_k$ sets): Each node of the SoC test scheduling graph is a set of cores that can be tested concurrently and no other core can be added to this set (we call these sets, power group sets or $N_k$ sets). For example for SoC of Table 1 with 5 cores, and given power characteristics ($P_m$), assuming that $P_{max}=300$, then $N_1$ can be {1, 2} and $N_2$ can be {1, 3, 5}.*

***Definition 2*** *($M_{SoC}$): The set of all $N_k$ sets in an SoC is called $M_{SoC}$. For example for the SoC of Table 1, $M_{SoC}$={{1, 2}, {1, 3, 5}, {1, 4}, {2, 3, 5}, {3, 4, 5}}.*

The algorithm of Figure 4 finds $M_{SoC}$ for an SoC. This algorithm is a recursive algorithm where inputs are the power characteristic for each core of the SoC, peak power upper bound of the SoC, and an $N_k$ set. First, the algorithm initializes $M_{SoC}$ with an empty set. Then, core $i$ with $P_{mi}$ is added to $N_k$, if the total power for $N_k$ does not exceed $P_{max}$. Then, the algorithm calls itself with a new peak power upper bound ($P_{max}-P_{mi}$) and $N_k$ (Lines 5). Finally, if there is no core in the SoC that can be added to $N_k$, the generated $N_k$ will be added to $M_{SoC}$.

***Definition 3*** *(incomplete sets): set of cores of an SoC where overall test power is lower than $P_{max}$, but there exist some other cores in SoC that can be tested with them, and yet the overall test power remains below the peak power ($P_{max}$). For example, {3, 5} is an incomplete set for the SoC of Table 1, because Core 1 can be added to this set and the overall power will not exceeds $P_{max}$.*

By defining $N_k$ sets, $M_{SoC}$, and incomplete sets, now we are able to define the test scheduling graph.

***Definition 4:*** *Test scheduling graph for an SoC is T = (G, E) in which G is the set of graph nodes. Each node of the graph corresponds to an $N_k$ set or an incomplete set. $E \subset G^2$ is the edges of the graph modeling the switching between two $N_k$ sets. Each Node k in the graph has a time that will be shown by $T_k$.*

We use test scheduling graph for modeling the concurrent testing of cores in a hybrid BIST architecture. For example Figure 5 depicts a test scheduling graph for applying all DTPs and PRTPs of cores with characteristic shown in Table 1.

```
Inputs: 1) core characteristics (SoC)
        2) Maximum power limit (P_max)
        3) A power group set (N_k)
Output: All N_k sets (M_SoC)

Finding M_SoC(core characteristics, P_max)
1  M_SoC={};
2  FOR each core i in SoC that is not in N_k DO
3    IF (P_mi<P_max) THEN
4      Include core i in N_k;
5      Finding M_SoC(core characteristics, P_max-P_mi, N_k);
6    END IF
7    IF (for each core i in SoC that is not in N_k,
       P_mi<P_max) THEN
8      Add N_k to M_SoC;
9    END IF
10 END FOR
   END
```

Figure 4. Algorithm for finding power group sets.

In each $N_k$ set, cores in the set are tested concurrently. When the *BIST* or the *Externaltest part* of a core, like $N_i$, in a group of cores is complete, that core is released from *BIST* or *External test part*. Each node is labeled by its time duration, i.e., $T_{node}$. $T_I$ corresponds to the time of the incomplete node *I*.

The test scheduling graph is performed considering power consumption of each core. Each node handles timing of *BIST* or *External test part* of the cores selected based on $N_k$. Selection between *BIST* or *External test part* of a core in a node depends on the core's $T(v_p)$ and $T(v_d)$, and will be discussed in the next section. This example only shows transitions from one node to another.

According to the test scheduling graph shown in Figure 5, first, the *BIST part* of Core 1 and the *External test part* of Core 2 are tested concurrently in Node 1. We move to Node 2 when the *BIST part* of Core 1 finishes ($T(v_p)$=0). In this case, the *BIST part* of Core 1 is released from Node 1 leaving 200 (400-200) of $T(v_d)$ of Core 2 to be handled by the next node. Then, the *External test part* of Core 2 and *BIST parts* of Core 3 and Core 5 are tested in Node 2. After 200 cycles, *External test part* of Core 2 and *BIST part* of Core 3 are released from Node 2, leaving 200 of

*BIST part* of Core 5. *BIST parts* of Core 2 and Core 5 and *External test part* of Core 3 are tested concurrently in Node 3. The *BIST part* of Core 5 is released from Node 3. In $I_1$ (the next node in the graph shown in Figure 5), the *BIST part* of Core 2 and *External test part* of Core 3 are tested concurrently. Note that nodes labeled by I refers to incompletesets, i.e., Definition 3, because $\{2, 3\} \notin M_{SoC}$.

*External test part* of Core 1 and *BIST part* of Core 4 are tested concurrently in Node 4 and the *External test part* of Core 1 is released from Node 4. Finally the remaining test of cores will be finished by $I_2$, $I_3$ and $I_4$. The total test time for this test scheduling graph is:

$$Time = \sum_{i \in G_T} T_i + \sum_i T_{I_i} \quad (4.1)$$

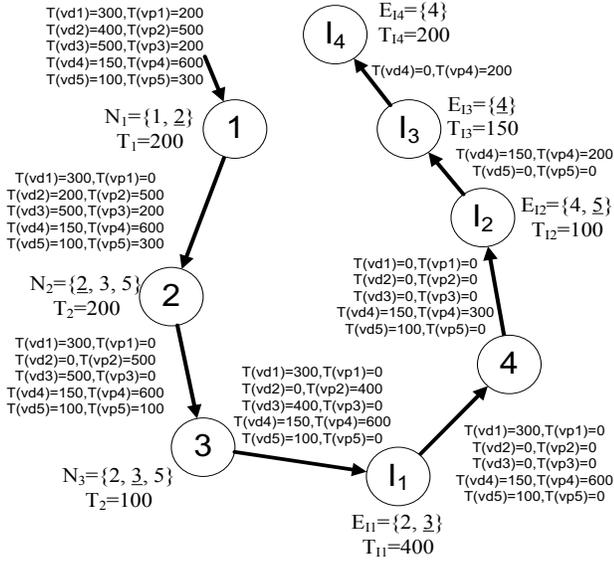

Figure 5. An example test scheduling graph.

According to the proposed architecture in Section III, we should have at most one *External test part* while using the functional bus for applying deterministic test patterns (cores tested externally are *underlined* in Figure 5). Because of high time penalty of pausing *External test parts* (due to state saving), we choose not to pause external tests while a core is being tested through TAM or functional bus. In the following sections, based on the proposed architecture and assumptions, we will discuss our algorithm to find an optimal test scheduling graph. We use this algorithm to reduce TAM as much as possible.

## V. ALGORITHM FOR TEST SCHEDULING

Figure 6 shows the proposed test scheduling algorithm based on test scheduling graph. The algorithm gets core characteristics and peak power consumption of an SoC. First, test generation according to algorithm shown in Figure 1 generates optimal DTPs and PRTPs for each core (Line 1). After that, the $M_{SoC}$ will be generated according to algorithm shown in Figure 4 (Line 2). The algorithm sorts all $N_k$ sets from $M_{SoC}$ by assigning a weight to them. This weight can help us decide which $N_k$ set should be selected first (Line 3 in Figure 6). Based on a weighted $N_k$ sets, i.e. *Weighted*_$M_{SoC}$ in Figure 6, the algorithm selects $N_k$ sets by starting from the highest weight, to make test scheduling graph,

TSG. This continues until all BIST parts and deterministic parts for all cores are covered (Line 4). After making the TSG, the algorithm adds some cores to incomplete nodes to be tested by their *BIST part* and generates new DTPs based on the added PRTPs in their BIST part (Lines 5-10). For example, consider again $I_1$ in Figure 5. As mentioned, $\{2, 3\}$ does not belong to $M_{SoC}$, but $\{2, 3, 5\}$ does. Then we can add *BIST part* to Core 5, by generating more PRTPs for Core 5, and including this BIST part to $I_1$. By adding extra PRTPs to Core 5 new DTPs should be generated. It is obvious that if we add some PRTPs to test a core, the generated number of deterministic test patterns decreases. After that, the algorithm updates the weighs of $N_k$ sets based on the newly generated DTPs and PRTPs. This process continues until exhausting all incomplete nodes.

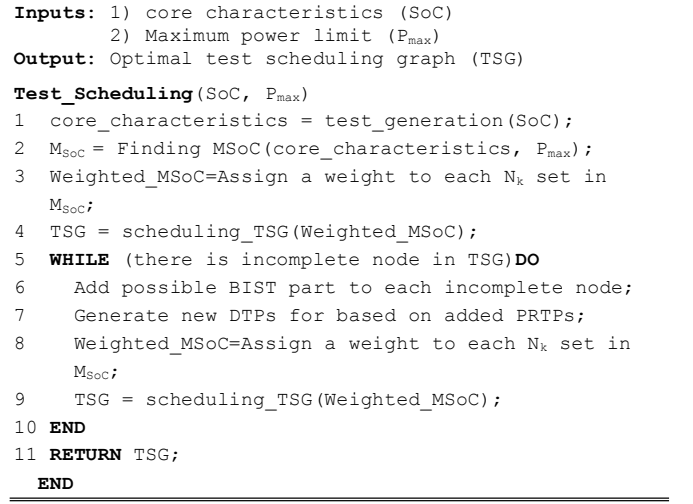

Figure 6. Proposed algorithm for power constraint test scheduling.

For giving a weight to $N_k$ sets ($W_{Nk}$), the following Heuristics are helpful:

**Heuristic 1.** To select $N_k$ sets for a test scheduling graph, sets of cores with longer *External test parts* are given a higher priority. So the weight for an $N_k$ set depends on the time of the longest *External test part* among the cores of the $N_k$ set, $\max_{i \in N_k} T(vd_i)$.

**Reasoning.** This is due to the fact that a pause for *External test parts* has more time overhead for global ATE that pausing *BIST parts* can be done.

**Heuristic 2.** As mentioned, to select an $N_k$ set, set of cores with the longest *External test part* is better to be selected first (Heuristic 1). The average of *BIST parts* of other cores in this set should be the largest of all available sets. So if we call $e_k$ the core with the longest *External test part* in $N_k$, the weight for $N_k$ set depends on the average of $T(v_{pe})$, such that $e \in N_k$ and $e \neq e_k$.

**Reasoning.** It can be observed that a combination of long *BIST parts* with long *External test part*, in an $N_k$ set can result in more concurrency. Then, the average of *BIST parts* in each $N_k$ set should be high.

Based on the above heuristics we have:

$$W_{N_k} \propto \max_{\substack{i \in N_k \\ i \neq e_k}} T(v_{di}) \quad (5.1)$$

$$W_{N_k} \propto \sum_{i \in N_k} \frac{T(v_{pi})}{|N_k| - 1} \quad (5.2)$$

$$W_{N_k} = \max_{i \in N_k} T(v_{di}) \times \sum_{\substack{i \in N_k \\ i \neq e_k}} \frac{T(v_{pi})}{|N_k| - 1} \quad (5.3)$$

The set with the highest weight will be selected as the node of the test scheduling graph. Within that node, the core with the highest value of $T(v_d)$ will be selected to be tested externally. After selecting an $N_k$, the weight of all remaining sets should be reevaluated according to the test parts covered in the selected $N_k$. If all *External test parts* of all cores are completed, or if it is not allowed to pause the *External test part* of the core that is being tested externally in the recently selected $N_k$ set, calculation of $W_{N_k}$ changes from that shown in Equation 5.3 to Equation 5.4 because in this case we do not have any *External test part*.

$$W_{N_k} = \sum_{i \in N_k} \frac{T(v_{pi})}{|N_k|} \quad (5.4)$$

## VI. EXPERIMENTAL RESULTS

For the experimental results, we used cores of $MCDS_1$ from [21] that is a version of d695 from ITC02 benchmark. The problem for comparing our results with available methods is availability of the gate level details of the cores. Then, d695 is the best choice because its cores are from ISCAS benchmarks. The characteristic of the SoC, i.e., power, frequency in each domain, etc., is exactly the same as $MCDS_1$ [21]. The ATALANTA test generator is used for determining DTPs for *External test part* of our hybrid BIST.

The optimal $N_{PRTP}$ and $N_{DTP}$ obtained by the proposed algorithm of Figure 1 for several d695 cores are shown in Table 2. This information will be used for initializing the SoC test scheduling process. We used ATALANTA for generating DTPs, an LFSR for generating PRTPs, and a parallel fault simulator. The clock frequency for *External test part* is 100MHz. PRTPs are applied in one clock cycle. Using scan chain for applying DTPs, the addition of primary inputs and pseudo primary inputs determine the number of clock cycles for applying each DTP. Finally, our proposed method test cycles is computed from Equation 6.1.

$$PMTC = (PMDV \times (PIs + PPIs) + (opt\_PRTP)) \quad (6.1)$$

In this equation, *PMDV* is the total number of selected deterministic test vectors, *opt_PRTP* is the number if pseudo random tests. PIs show the number of primary inputs and PPIs are number of pseudo primary inputs of the circuit.

Table 2. Optimal number of applying PRTPs and DTPs

| cores | $S_B/S_E$ | $N_{DTP}$ phase1 | $N_{PRTP}$ PRTPs | $N_{DTP}$ phase2 | Test time (µs) |
|---|---|---|---|---|---|
| C6288 | 32 | 2 | 42 | 0 | 1.06 |
| C7552 | 207 | 10 | 2482 | 94 | 240.1 |
| S838 | 67 | 23 | 357 | 38 | 44.44 |
| S5378 | 214 | 10 | 1953 | 73 | 197.15 |
| S9234 | 247 | 58 | 3960 | 127 | 496.55 |
| S13207 | 700 | 89 | 8934 | 48 | 1048.34 |
| S15850 | 611 | 115 | 959 | 166 | 1726.5 |
| S35932 | 1763 | 6 | 225 | 0 | 108.03 |
| S38417 | 1664 | 162 | 9834 | 353 | 8667.94 |
| S38584 | 1464 | 55 | 4573 | 180 | 3486.13 |

The results of the obtained number of the sets for $M_{SoC}$ for some SoCs according to the algorithm of Figure 4 are shown in Table 3. We used d695 benchmark, $MCDS_1$ from [21], and hCAD01 from [20] to show the number of sets and CPU time of the proposed algorithm.

Table 3. Number of sets of $M_{SoC}$/CPU Time

| | TAM | $P_{max}$=1500 | $P_{max}$=2000 | $P_{max}$=3000 |
|---|---|---|---|---|
| |$M_{d695}$| | Un Lim | 29/1ms< | 54/2ms | 126/8ms |
| | 32pin | 10/1ms< | 8/1ms | 8/1ms |
| | 64pin | 29/1ms< | 43/2ms | 29/3ms |
| | 128pin | 29/2ms | 54/3ms | 125/8ms |
| |$M_{MCDS}$| | Un Lim | 88/4ms | 227/9ms | 767/81ms |
| | 32pin | 17/2ms | 13/1ms | 13/2ms |
| | 64pin | 64/3ms | 101/5ms | 96/8ms |
| | 128pin | 88/5ms | 227/11ms | 669/60ms |
| | TAM | $P_{max}$=3000 | $P_{max}$=4500 | Un Lim |
| |$M_{hCADT01}$| | Un Lim | 6/1ms | 11/1ms | 1/1ms< |
| | 32pin | 6/1ms | 13/1ms | 19/2ms |
| | 64pin | 6/1ms | 11/1ms | 1/1ms< |
| | 128pin | 6/1ms | 11/1ms | 1/1ms< |

The results of TAT for the proposed method are shown in Table 4 and Table 5, while using a fixed clock for ATE and different peak power and TAM width limitations. For comparison, there was a miss match between the number of generated test patterns by ATALANTA for full coverage and the number of test patterns reported in ITC02 benchmarks for each core. For example, for core "c6288" (that is the first core of "d695") ATALANTA generates 33 DTPs for full coverage, while 12 TPs are reported in ITC02 benchmark. So, we implemented the proposed method in [21] with the new number of test patterns generated by ATALANTA for full coverage.

The results are categorized by different peak power limitation, different number of TAM sizes, and fixed or flexible number of scan chains for each core. The fixed number of scan chains wereobtained from those reported in the ITC02 benchmark. In the flexible number of scan chains, no limitation is considered for determining the number of scan chains for each core. As shown, a considerable TAT reduction is obtained by the proposed method in comparison with [21] and [22].

## VII. CONCLUSIONS ANDFUTUREWORKS

In this paper, a concurrent Hybrid BIST method for reducing SoC test time is proposed. The most important constraint is the power limitation of the chip. An algorithm to find the most suitable cores to be tested concurrently is proposed. During the test process, applying Deterministic Test Patterns and Pseudo Random Test Patterns can be done together to reduce the test time. Experimental results show that using this method provides a considerable reduction in test application time compared to the previous methods for power constrained testing. This process can be applied for cycle accurate power model that improves the TAT.

Table 4. Test application time of Proposed Method (PM (*μs*)) for fix number of scan chain in comparison with [21] and [22] for multi-clock domain SoCMCDS$_1$.

| $F_{ATE}$ | $P_{MAX}$ | TAM width | | | | | | | | | | | |
|---|---|---|---|---|---|---|---|---|---|---|---|---|---|
| | | 32 pin | | | | 64 pin | | | | 128 pin | | | |
| | | TAT ([21]) | TAT ([22]) | PM | | TAT ([21]) | TAT ([22]) | PM | | TAT ([21]) | TAT ([22]) | PM | |
| | | | | TAT | CPU Time | | | TAT | CPU Time | | | TAT | CPU Time |
| **200MHz** | 1500 | 2500.09 | 624.29 | 474.4604 | 482 | 2037.076 | 621.92 | 472.65 | 510 | 2037.076 | 620.34 | 473.4 | 415 |
| | 2000 | 1886.68 | 622.5 | 473.1 | 569 | 1615.59 | 620.125 | 471.29 | 552 | 1615.59 | 617.28 | 470.1 | 521 |
| | 2500 | 1879.49 | 621.23 | 472.1348 | 727 | 1358.592 | 618.86 | 470.33 | 724 | 1358.592 | 614.73 | 467.1 | 407 |
| **100MHz** | 1500 | - | - | - | - | 2500.09 | 1243.84 | 945.31 | 747 | 2037.076 | 1240.68 | 942.9 | 509 |
| | 2000 | - | - | - | - | 1886.682 | 1240.2 | 942.55 | 498 | 1615.59 | 1234.56 | 938.26 | 509 |
| | 2500 | - | - | - | - | 1879.49 | 1237.2 | 940.27 | 391 | 1359.172 | 1229.46 | 937.36 | 353 |
| **50MHz** | 1500 | - | - | - | - | - | - | - | - | 2500.09 | 1740.5 | 1422.7 | 516 |
| | 2000 | - | - | - | - | - | - | - | - | 1886.682 | 1734.5 | 1320 | 478 |
| | 2500 | - | - | - | - | - | - | - | - | 1879.49 | 1729 | 1314 | 521 |

Table 5. Test application time of Proposed Method (PM (*μs*)) for flexible number of scan chain in comparison with [22] for multi-clock domain SoCMCDS$_1$.

| $F_{ATE}$ | $P_{MAX}$ | TAM width | | | | | | | | |
|---|---|---|---|---|---|---|---|---|---|---|
| | | 32 pin | | | 64 pin | | | 128 pin | | |
| | | TAT ([22]) | PM | | TAT ([22]) | PM | | TAT ([22]) | PM | |
| | | | TAT | CPU Time | | TAT | CPU Time | | TAT | CPU Time |
| **200MHz** | 1500 | 453 | 345.28 | 398 | 231 | 185.56 | 392 | 120 | 95.2 | 394 |
| | 2000 | 453 | 345.28 | 435 | 231 | 185.56 | 304 | 120 | 95.2 | 282 |
| | 2500 | 453 | 345.28 | 392 | 231 | 185.56 | 597 | 120 | 95.2 | 599 |
| **100MHz** | 1500 | 907.12 | 689.41 | 361 | 463 | 341.88 | 396 | 241 | 183 | 390 |
| | 2000 | 907.12 | 689.41 | 326 | 463 | 341.88 | 393 | 241 | 183 | 392 |
| | 2500 | 907.12 | 689.41 | 491 | 463 | 341.88 | 264 | 241 | 183 | 398 |
| **50MHz** | 1500 | 1814 | 1388.64 | 553 | 927 | 714.52 | 704 | 482 | 376 | 384 |
| | 2000 | 1814 | 1388.64 | 394 | 927 | 714.52 | 396 | 482 | 376 | 672 |
| | 2500 | 1814 | 1388.64 | 393 | 927 | 714.52 | 499 | 482 | 376 | 385 |